\documentclass[twocolumn,lineno,dvipsnames, usenames]{aastex63}
\usepackage{aas_macros}
\usepackage{graphicx}
\usepackage{dcolumn}
\usepackage{bm}
\usepackage{amsmath}
\usepackage{float}
\usepackage{lipsum}
\usepackage{xcolor}
\usepackage{textcomp}
\usepackage{latexsym}
\usepackage{mathtools}
\usepackage{url}
\usepackage{comment}
\usepackage[utf8]{inputenc}
\usepackage[normalem]{ulem}
\usepackage{xspace}
\usepackage{acronym}

\newcommand{\chieff}{\ensuremath{\chi_{\mathrm{eff}}}\xspace}

\newcommand{\LIGOlabMIT}{\affiliation{LIGO Laboratory, Massachusetts Institute of Technology, 185 Albany St, Cambridge, MA 02139, USA}}
\newcommand{\MKI}{\affiliation{Department of Physics and Kavli Institute for Astrophysics and Space Research, Massachusetts Institute of Technology, 77 Massachusetts Ave, Cambridge, MA 02139, USA}}
\newcommand{\CCA}{\affiliation{Center for Computational Astrophysics, Flatiron Institute, 162 Fifth Avenue, New York, NY 10010, USA}}
\newcommand{\SUNY}{\affiliation{Department of Physics and Astronomy, Stony Brook University, Stony Brook, NY 11794, USA}}

\begin{document}
\title{The binary black hole spin distribution likely broadens with redshift}
\correspondingauthor{Sylvia Biscoveanu}
 \email{sbisco@mit.edu}
\author[0000-0001-7616-7366]{Sylvia Biscoveanu} \MKI \LIGOlabMIT
\author[0000-0001-9892-177X]{Thomas A. Callister} \CCA
\author[0000-0001-8040-9807]{Carl-Johan Haster} \MKI \LIGOlabMIT
\author[0000-0003-3896-2259]{Ken K.Y. Ng} \MKI \LIGOlabMIT
\author[0000-0003-2700-0767]{Salvatore Vitale} \MKI \LIGOlabMIT
\author[0000-0003-1540-8562]{Will M. Farr} \CCA \SUNY

\begin{abstract}
    The population-level distributions of the masses, spins, and redshifts of binary black holes (BBHs) observed using gravitational waves can shed light on how these systems form and evolve. Because of the complex astrophysical processes shaping the inferred BBH population, models allowing for correlations among these parameters will be necessary to fully characterize these sources. We hierarchically analyze the BBH population detected by LIGO and Virgo with a model allowing for correlations between the effective aligned spin and the primary mass and redshift. We find that the width of the effective spin distribution grows with redshift at 98.6\% credibility. We determine this trend to be robust under the application of several alternative models and additionally verify that such a correlation is unlikely to be spuriously introduced using a simulated population. We discuss the possibility that this correlation could be due to a change in the natal black hole spin distribution with redshift.
\end{abstract}

\section{Introduction}
The growing catalog of compact binary mergers detected with gravitational
waves~\citep{LIGOScientific:2018mvr, LIGOScientific:2021usb, LIGOScientific:2021djp} has allowed for increasingly precise
characterization of the population properties of binary black holes (BBHs) and
neutron stars~\citep{LIGOScientific:2021psn}, which can shed light on how these
systems form and evolve~\citep{Wong:2020jdt, Zevin:2020gbd, Bouffanais:2021wcr}.
The latest data from the Advanced LIGO~\citep{TheLIGOScientific:2014jea} and
Virgo~\citep{TheVirgo:2014hva} observatories has revealed that the BBH mass
distribution has substructure~\citep{Tiwari:2020otp, Edelman:2021zkw,
Li:2021ukd, Veske:2021qis} beyond a smooth power-law and single peak or break at
$\sim 40~\mathrm{M}_{\odot}$~\citep{Fishbach:2017zga, Talbot:2018cva,
LIGOScientific:2018jsj}. BBH spins are found to be small but
non-zero~\citep{Wysocki:2018, Roulet:2018jbe,Miller:2020zox,Garcia-Bellido:2020pwq, Biscoveanu:2020are},
with some of their tilts misaligned to the orbital angular
momentum~\citep{Talbot:2017yur, LIGOScientific:2018jsj, LIGOScientific:2021psn}, although these conclusions have been challenged when applying a different population
model~\citep{Roulet:2021hcu, Galaudage:2021rkt}. The
merger rate is found to evolve with redshift at a rate consistent with the star
formation rate~\citep{Madau:2014bja, Fishbach_2018, LIGOScientific:2021psn}.

In addition to fitting the mass, spin, and redshift distributions independently,
previous works have looked for correlations among these
parameters~\citep{Safarzadeh:2020mlb, Callister:2021fpo, LIGOScientific:2021psn, Tiwari:2021yvr,
Franciolini:2022iaa}. Such correlations can be imprinted via evolutionary processes within a single formation channel or can be caused by the presence of multiple populations arising from distinct formation channels. For example,
systems formed dynamically in dense environments via repeated mergers are
expected to be more massive and have higher
spins~\citep{PortegiesZwart:2002iks,Benacquista:2011kv,Miller:2008yw,McKernan:2012rf,
Antonini:2016gqe, Rodriguez:2015oxa, Gerosa:2017kvu, Fishbach:2017dwv,
Rodriguez:2019huv, Kimball:2020opk, Gerosa:2021mno}. \cite{Callister:2021fpo}
and \cite{LIGOScientific:2021psn} found statistically significant evidence for a
correlation between the distribution of the mass-weighted spin aligned with the
orbital angular momentum, \chieff, and the binary mass ratio, $q$, where the
mean of the \chieff distribution increases for more extreme mass ratios. This
sort of correlation is unexpected for most individual BBH formation models, with
the possible exception of formation in the disks of active galactic nuclei~\citep{McKernan:2012rf, Stone:2016wzz, Mckernan:2017ssq,
McKernan:2019beu, Tagawa:2019osr, Callister:2021fpo} and super-Eddington accretion during stable mass transfer for systems formed via isolated binary evolution~\citep{Bavera:2020uch, Zevin:2022wrw}, so may hint at the
superposition of multiple populations formed via independent channels with
unique $q$ vs. \chieff signatures. 

\citet{Safarzadeh:2020mlb} examined whether the effective spin distribution
correlates with various mass parameters of the binary.  They found a possible
negative correlation between the mean effective spin and each of the chirp mass ($\mathcal{M} = (m_{1}m_{2})^{3/5}/(m_1 + m_2)^{1/5}$), total mass, and primary mass parameters and a possible
positive correlation between the dispersion of the effective spin and mass;
neither finding reached high statistical significance ($\sim 80\%$ credibility depending on the mass and correlation parameters).

\cite{LIGOScientific:2021psn} and \cite{Tiwari:2021yvr} also found that the spread in the component spins aligned with the orbital angular momentum may increase with the binary chirp mass. This is explained as an
effect of the paucity of events at large chirp masses, which manifests as a
degradation of the constraint on the spin distribution, rather than a firm
measurement of an increase in the width of the distribution at larger chirp
masses. \cite{Franciolini:2022iaa} also find initial evidence of an evolution of
the \chieff distribution with increasing total mass, which they interpret as evidence for a sub-population of dynamically-formed or primordial black hole binaries~\citep{DeLuca:2020bjf, DeLuca:2020fpg, DeLuca:2020qqa,
Franciolini:2021xbq}, as both models predict correlations between mass and spin.

Previous works have also explored potential correlations between the BBH mass and redshift distributions~\citep{Fishbach:2021yvy, LIGOScientific:2021psn}. Heavier black holes are predicted to form from lower-metallicity stellar progenitors at higher redshifts, which could lead to such a correlation~\citep{Belczynski:2017gds, Mapelli:2019bnp, Neijssel:2019irh}. For systems formed via isolated binary evolution, those that undergo a common envelope event tend to be both less massive and have shorter delay times, merging at higher redshifts~\citep{vanSon:2021zpk}. Motivated by these theoretical predictions, \cite{Fishbach:2021yvy} find that redshift dependence of the maximum black hole mass is required if the primary mass distribution has a sharp cutoff, but a gradual tapering of the primary mass distribution is consistent with no evolution with redshift, a finding which was confirmed with the latest catalog of events by \cite{LIGOScientific:2021psn}.

In this work, we search for correlations between the \chieff distribution and
the primary mass and redshift distributions. We find robust evidence of a
correlation between \chieff and redshift, where the width of the \chieff
distribution increases with redshift. We also find a weaker correlation between
the width of the \chieff distribution and primary mass. When allowing the
\chieff distribution to correlate with both redshift and primary mass, we find a
preference in the data for a correlation with one or the other, although we
cannot distinguish which. In Section~\ref{sec:methods}, we describe the
Bayesian methods and models employed in our analysis. The results on data
from the latest GWTC-3 catalog~\citep{LIGOScientific:2021djp} of compact binaries observed by LIGO-Virgo are presented in Section~\ref{sec:results}.
Section~\ref{sec:validation} includes a validation of our results using
simulated populations and alternative models applied to the real data. We
conclude and comment on the potential astrophysical implications of our finding in
Section~\ref{sec:conclusions}.

\section{Methods}
\label{sec:methods}
We employ the framework of hierarchical Bayesian inference to constrain the hyper-parameters governing the population-level distributions of the masses, spins, and redshifts of the binary black hole systems detected by LIGO and Virgo. We assume the distribution of primary masses, $m_1$, is described by the sum of a truncated power law with low-mass smoothing and a Gaussian component~\citep{Talbot:2018cva}, dubbed the \textsc{Power Law + Peak} model in \cite{LIGOScientific:2020kqk, LIGOScientific:2021psn}, and that the mass ratio distribution is also a power law, bounded such that the minimum and maximum masses of the secondary component are the same as those of the primary~\citep{Fishbach:2019bbm}. The merger rate is allowed to evolve with redshift as a power-law, following the model in \citet{Fishbach_2018} and \citet{LIGOScientific:2021psn}. The hyper-parameters governing these mass and redshift distributions are described in Table~\ref{tab:mass_z_hyper_param_priors}. 

\begin{table*}
    \centering
\begin{tabular}{|p{2cm} ||p{5.5cm} p{3cm}|}
    \hline
    Parameter & Description & Prior\\
    \hline
    $\alpha$ & $m_{1}$ power-law index & U(-4, 12) \\
    $\beta$  & $q$ power-law index & U(-4, 7) \\
    $m_{\max}$ & maximum BH mass & $\mathrm{U}(30~\mathrm{M}_{\odot}, 100~\mathrm{M}_{\odot})$\\
    $m_{\min}$  & minimum BH mass & $\mathrm{U}(2~\mathrm{M}_{\odot}, 10~\mathrm{M}_{\odot})$ \\
    $\delta_{m}$ & low-mass smoothing parameter & $\mathrm{U}(0~\mathrm{M}_{\odot}, 10~\mathrm{M}_{\odot})$\\
    $\mu_{m}$ & PISN peak location & $\mathrm{U}(20~\mathrm{M}_{\odot}, 50~\mathrm{M}_{\odot})$ \\
    $\sigma_{m}$ & PISN peak width & $\mathrm{U}(1~\mathrm{M}_{\odot}, 10~\mathrm{M}_{\odot})$ \\
    $\lambda$ & fraction of systems in PISN peak & $\mathrm{U}(0, 1)$\\
    $\lambda_{z}$ & $z$ power-law index & $\mathrm{U}(-2, 10)$\\
    \hline
\end{tabular}
\caption{Mass and redshift hyper-parameter priors for the standard \textsc{power-law + peak} and \textsc{power-law redshift} distributions}
\label{tab:mass_z_hyper_param_priors}
\end{table*}

Our goal is to determine if there is a correlation between the BBH spin distribution and the distributions of masses or redshifts. To this end, we modify the correlated spin model introduced in \citet{Callister:2021fpo} so that the effective aligned spin~\citep{Damour:2001, Ajith:2009bn, Ajith:2011ec, Santamaria:2010yb}, 
\begin{align}
    \chieff = \frac{\chi_{1}\cos{\theta_{1}} + q\chi_{2}\cos{\theta_{2}}}{1 + q},
    \label{eq:chi_eff}
\end{align}
is modeled as a truncated Gaussian on $[-1, 1]$ whose mean and variance can each evolve linearly with primary mass and redshift, 
\begin{align}
    \label{eq:chi_eff_pop}
    \pi_{\mathrm{pop}}(\chieff &| \Lambda_{\chi_\mathrm{eff}}, m_1, z) = \mathcal{N}(\chieff; \mu_{\chi}, \sigma_{\chi}),\\
    \mu_{\chi}(z, m_{1}) &= \mu_{0} + \delta\mu_z (z - 0.5) + \delta\mu_{m_{1}}\left(\frac{m_{1}}{10~\mathrm{M}_{\odot}} -1\right)\\
    \log{\sigma_{\chi}}(z, m_{1}) &= \log{\sigma_0} + \delta\hspace{-0.5mm}\log{\sigma_z}(z - 0.5) \\ \nonumber &+ \delta\hspace{-0.5mm}\log{\sigma_{m_{1}}}\left(\frac{m_{1}}{10~\mathrm{M}_{\odot}} -1\right),
\end{align}
where we use $\log$ to indicate log base 10 everywhere and $\ln$ to indicate the natural logarithm.
The pivot points of $z=0.5$ and $m_1=10~\mathrm{M}_{\odot}$ are chosen to be near the peak of the population distributions of those parameters, but we have verified that the exact values do not affect our results. To avoid unphysically narrow distributions using the model above, we impose a cut such that
\begin{align}
    \pi_{\mathrm{pop}}(\chieff &| \Lambda_{\chi_\mathrm{eff}}) = \begin{cases}
    0, \quad \mathrm{if\,}\log_{10}\sigma_{\chi}(z, m_{1}) < -3\\
    \mathcal{N}(\chieff; \mu_{\chi}, \sigma_{\chi}), \quad \mathrm{elsewhere}.
    \end{cases}
\end{align}
The full set of $\Lambda_{\chieff}$ parameters are described in Table~\ref{tab:spin_hyper_param_priors}. We note that unlike the model of \cite{Callister:2021fpo}, our model in Eq.~\ref{eq:chi_eff_pop} does not allow for correlations between \chieff and mass ratio. However, we amend the model to incorporate mass ratio correlations later in Section~\ref{sec:results}.
\begin{table*}
    \centering
\begin{tabular}{|p{2cm} ||p{5.5cm} p{3cm}|}
    \hline
    Parameter & Description & Prior\\
    \hline
    $\mu_{0}$ & Independent $\chieff$ mean & $\mathrm{U}(-1, 1)$\\
    $\log\sigma_{0}$ & Log of independent $\chieff$ width & $\mathrm{U}(-1.5, 0.5)$ \\
    $\delta\mu_{q}$ & $q$-dependent $\chieff$ mean & $\mathrm{U}(-2.5, 1)$\\
    $\delta\hspace{-0.5mm}\log\sigma_{q}$ & Log of $q$-dependent $\chieff$ width & $\mathrm{U}(-2, 1.5)$ \\
    $\delta\mu_{z}$ & $z$-dependent $\chieff$ mean & $\mathrm{U}(-2.5, 1)$\\
    $\delta\hspace{-0.5mm}\log\sigma_{z}$ & Log of $z$-dependent $\chieff$ width & $\mathrm{U}(-0.5, 1.5)$ \\
    $\delta\mu_{m_1}$ & $m_{1}$-dependent $\chieff$ mean & $\mathrm{U}(-2.5, 1)$\\
    $\delta\hspace{-0.5mm}\log\sigma_{m_1}$ & Log of $m_{1}$-dependent $\chieff$ width & $\mathrm{U}(-2, 1.5)$ \\
    \hline
\end{tabular}
\caption{Spin hyper-parameter priors and descriptions for the model in  Eq.~\ref{eq:chi_eff_pop}}
\label{tab:spin_hyper_param_priors}
\end{table*}

Given a set of posterior samples for the binary parameters $\theta$ of $N$ individual BBH events, they can be combined to obtain posteriors on the hyper-parameters governing the population-level distributions:
\begin{align}
    \label{eq:hyper-posterior}
    p(\Lambda | \{d\}) &\propto \mathcal{L}(\{d\} | \Lambda)\pi(\Lambda),\\
    \label{eq:hyper-likelihood}
    \mathcal{L}(\{d\} | \Lambda) &= \frac{1}{\alpha(\Lambda)^{N}} \prod_{i}^{N}\sum_{j}\frac{\pi_{\mathrm{pop}}( \theta_{i,j} | \Lambda)}{\pi_{\mathrm{PE}}(\theta_{i,j})}.
\end{align}
The factor of $\pi(\Lambda)$ in Eq.~\ref{eq:hyper-posterior} represents the prior probability for the hyper-parameters, given in Tables~\ref{tab:mass_z_hyper_param_priors}-\ref{tab:spin_hyper_param_priors}. The likelihood in Eq.~\ref{eq:hyper-likelihood} consists of a Monte Carlo integral over the posterior samples $j$ for each individual event $i$, where $\pi_{\mathrm{pop}}(\theta_{i,j} | \Lambda)$ is the product of the population distributions for the masses, redshift, and $\chi_\mathrm{eff}$ in Eq.~\ref{eq:chi_eff_pop}. $\pi_\mathrm{PE}(\theta_{i,j})$ is the original prior that was applied during the parameter estimation (PE) of the binary parameters for individual events. Finally, the factor of $\alpha(\Lambda)^{N}$ accounts for the fact that the observed BBH sources are a biased sample of the underlying astrophysical distribution~\citep{Loredo_2004, Thrane:2018qnx, Mandel_2019, Vitale:2020aaz}. This selection effect arises from the dependence of the sensitivity of the detector network on the intrinsic parameters of the source. 

We use the \textsc{GWPopulation} package~\citep{Talbot:2019} and the \textsc{dynesty} nested sampler~\citep{Speagle_2020} to obtain samples from the hyper-parameter posterior in Eq.~\ref{eq:hyper-posterior}. We include the 69 BBH events reported in GWTC-3 with a false alarm rate (FAR) less than 1 per year~\citep{LIGOScientific:2021djp}, as was done for the BBH analyses in \cite{LIGOScientific:2021psn}. We use the individual-event posterior samples publicly released by LIGO and Virgo~\citep{samples_gwtc1, data_release_samples, ligo_scientific_collaboration_and_virgo_2021_5117703, ligo_scientific_collaboration_and_virgo_2021_5546663} obtained with the IMRPhenomPv2 waveform model for events first published in GWTC-1~\citep{LIGOScientific:2018mvr}, and using a combination of different waveform models including the effects of spin precession and higher-order modes for events reported in later catalogs\footnote{These correspond to the \texttt{PublicationSamples}, \texttt{PrecessingSpinIMRHM}, and \texttt{C01:Mixed} datasets for GWTC-2, GWTC-2.1, and GWTC-3 events, respectively.}~\citep{LIGOScientific:2020ibl, LIGOScientific:2021usb, LIGOScientific:2021djp}. We calculate $\alpha(\Lambda)$ following the method described in \cite{Farr_2019}, using the sensitivity estimates for BBH systems released at the end of the most recent LIGO-Virgo observing run (O3) obtained via a simulated injection campaign~\citep{ligo_scientific_collaboration_and_virgo_2021_5636816}. 

\begin{figure*}
\centering
\includegraphics[width=\textwidth]{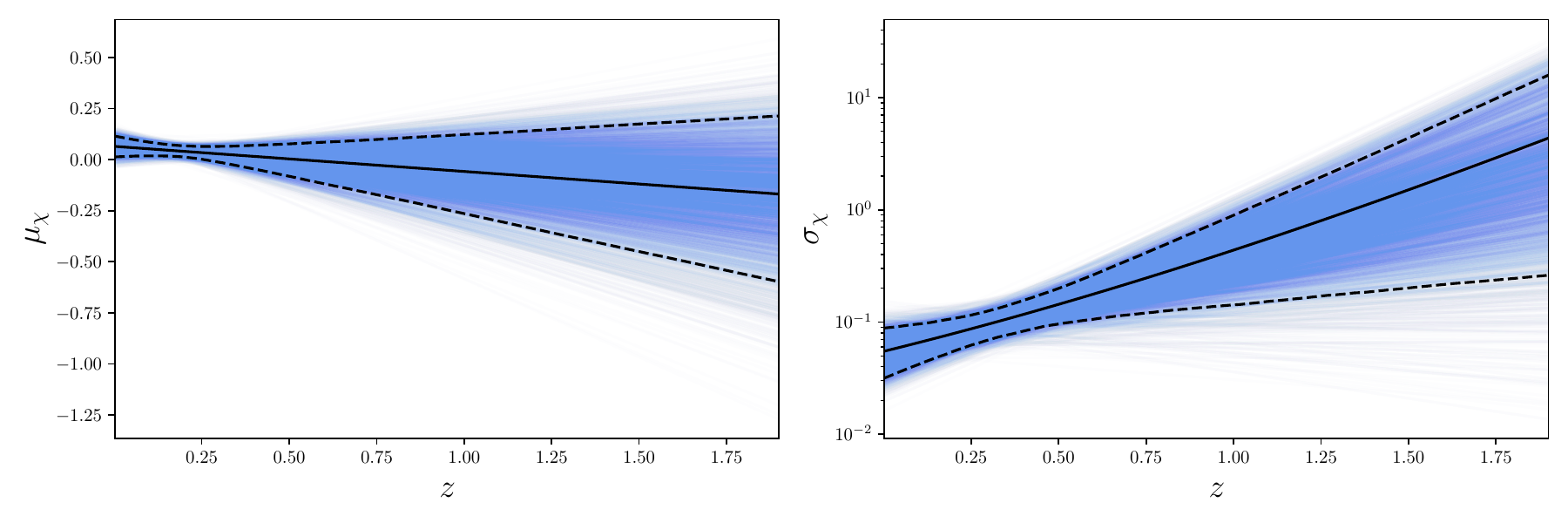}
\caption{Posteriors for the mean, $\mu_{\chi}$, and standard deviation, $\sigma_{\chi}$, of the binary black hole effective spin distribution as a function of redshift obtained for GWTC-3 events. The solid black line shows the mean, while the dashed black lines show the 90\% credible region.}
\label{fig:mu_sigma_scaling_z}
\end{figure*}

\section{Results for GWTC-3}
\label{sec:results}
\begin{figure}
\centering
\includegraphics[width=\columnwidth]{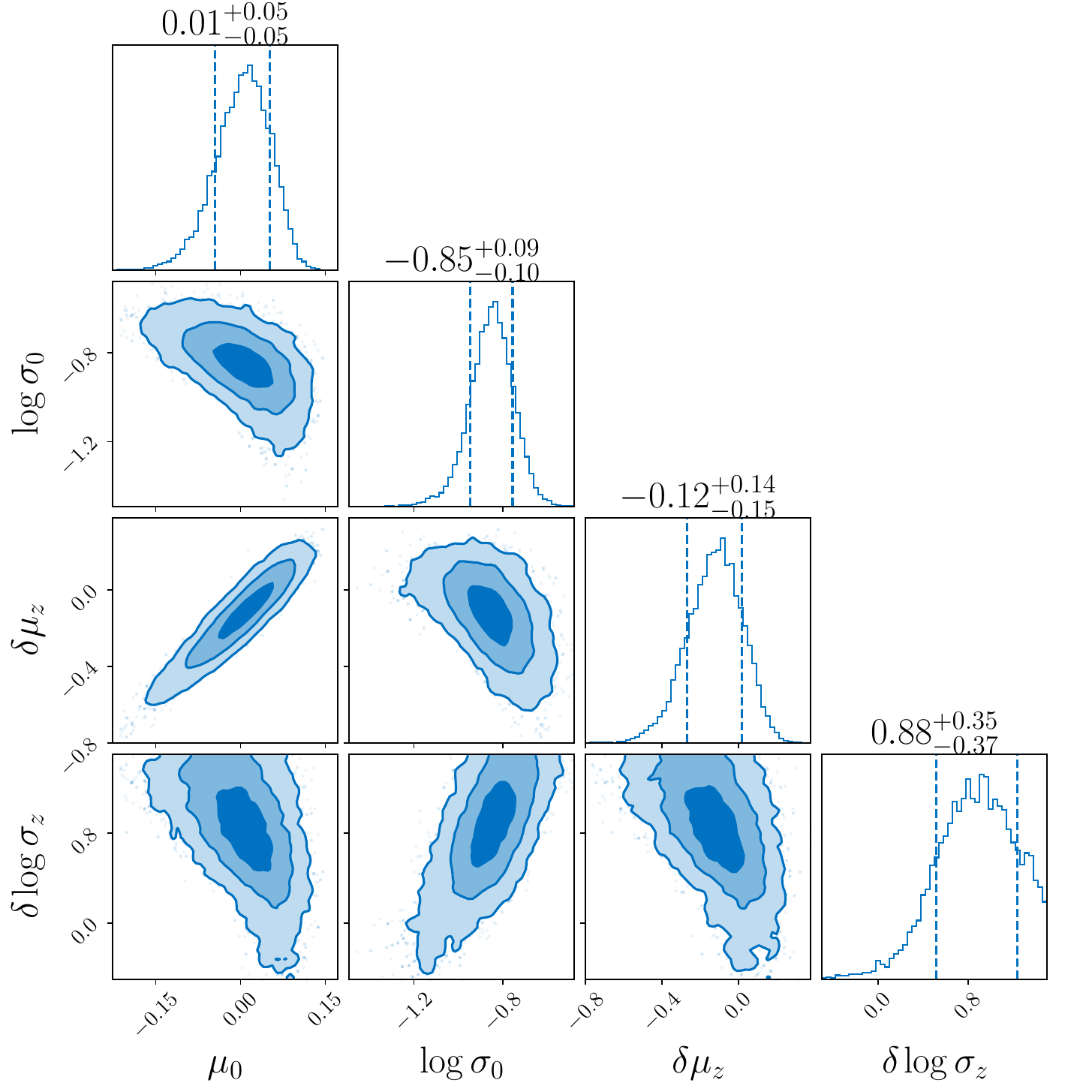}
\caption{Spin hyper-parameter posteriors obtained for GWTC-3 events when allowing for redshift correlations only. The colors indicate the 1, 2, and $3\sigma$ 2D credible regions, the dashed lines on the individual histograms show the 1D $1\sigma$ credible interval, and the median and $1\sigma$ credible interval are printed above each histogram.}
\label{fig:spin_corner_z}
\end{figure}

\begin{figure}
\centering
\includegraphics[width=\columnwidth]{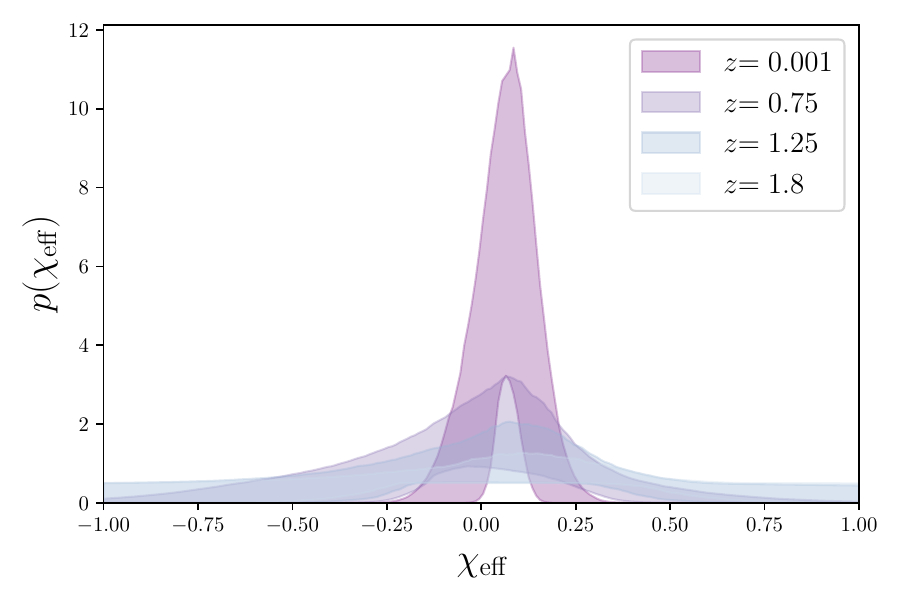}
\caption{90\% credible region for slices of the \chieff distribution at four different values of redshift.}
\label{fig:slices_z}
\end{figure}

We first allow the \chieff distribution to be correlated with redshift and primary mass individually. We recover mass and redshift hyper-parameter posteriors consistent with those reported in \cite{LIGOScientific:2021psn}. The evolution of $\mu_{\chi}$ and $\sigma_{\chi}$ as a function of redshift for individual hyper-parameter posterior samples obtained with the model only allowing for redshift correlations is shown in Fig.~\ref{fig:mu_sigma_scaling_z}. In this case the $\delta\mu_{m_1}$ and $\delta\hspace{-0.5mm}\log\sigma_{m_1}$ parameters are fixed to zero. While $\mu_{\chi}$ does not exhibit significant evolution with redshift, $\sigma_{\chi}$ increases with redshift. The corresponding posteriors on $\Lambda_{\chi_\mathrm{eff}}$ are shown in Fig.~\ref{fig:spin_corner_z}. The $\delta\mu_{z}$ posterior is consistent with 0, indicating no evolution of the mean of the \chieff distribution as a function of redshift, but we find that $\delta\hspace{-0.5mm}\log\sigma_{z}=0$ is disfavored at 98.6\% credibility. We recover $\delta\hspace{-0.5mm}\log\sigma_{z}=0.93^{+0.54}_{-0.54}$, (maximum posterior value and 90\% credible interval calculated with the maximum posterior density method) indicating that the spin distribution broadens with increasing redshift. This correlation can also be visualized by examining slices of the \chieff distribution at fixed redshift, as shown in Fig.~\ref{fig:slices_z}. The thickness of the 90\% credible region for each slice is comparable, but the distribution becomes noticeably broader with increasing redshift. This indicates that the apparent correlation between the width of the spin distribution and the redshift is not dominated by increased uncertainty in the constraint on the distribution at higher redshifts. The location of the peak of the \chieff distribution does not vary significantly between the different redshift slices, consistent with the lack of observed evolution in $\mu_{\chi}$.

\begin{figure}
\centering
\includegraphics[width=\columnwidth]{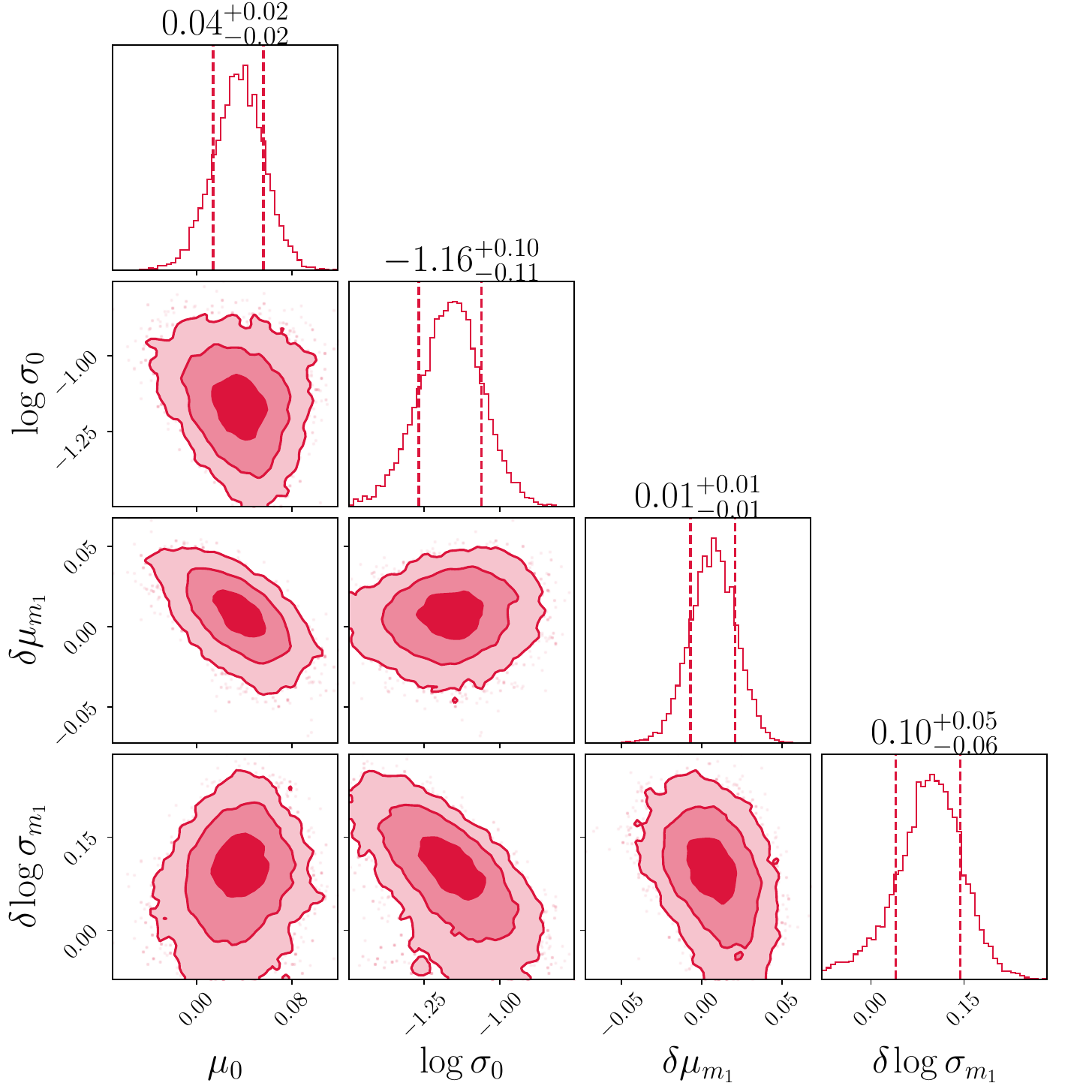}
\caption{Spin hyper-parameter posteriors obtained for GWTC-3 events when allowing for a correlation between \chieff and primary mass only. The colors indicate the 1, 2, and $3\sigma$ 2D credible regions, the dashed lines on the individual histograms show the 1D $1\sigma$ credible interval, and the median and $1\sigma$ credible interval are printed above each histogram.}
\label{fig:primary_mass_corner}
\end{figure}

\begin{figure}
\centering
\includegraphics[width=\columnwidth]{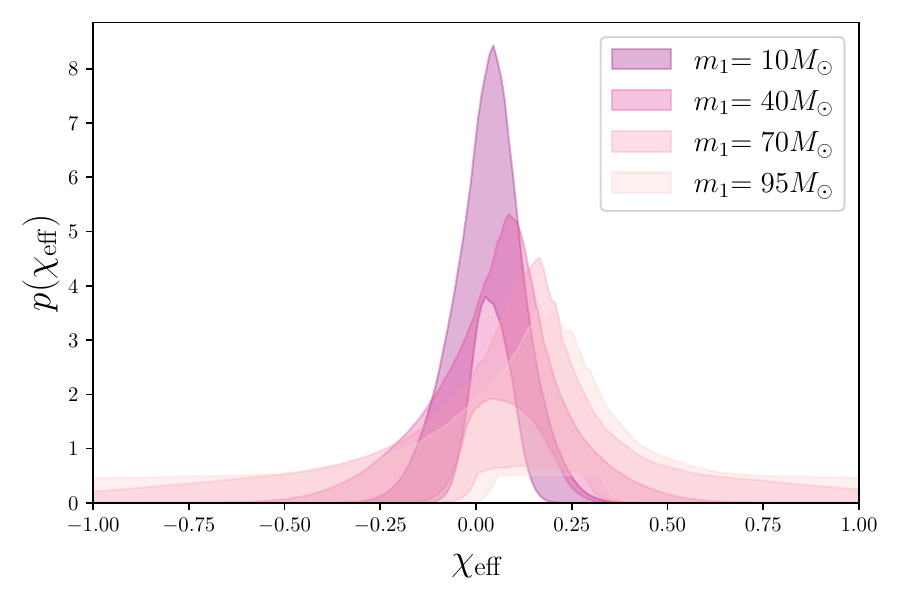}
\caption{90\% credible region for slices of the \chieff distribution at four different values of primary mass.}
\label{fig:primary_mass_slices}
\end{figure}

While these results present compelling evidence for a correlation between \chieff and redshift, it is possible that the BBH mass and redshift distributions are themselves correlated, so that our recovered redshift correlation is actually a manifestation of a mass vs. \chieff correlation viewed through the wrong model. To test this, we now allow the \chieff distribution to be correlated only with primary mass, fixing $\delta\mu_z=0,\  \delta\hspace{-0.5mm}\log\sigma_z=0$. The posteriors for the spin hyper-parameters for the primary mass correlation model are shown in Fig.~\ref{fig:primary_mass_corner}, and the distributions for \chieff along different slices in primary mass are shown in Fig.~\ref{fig:primary_mass_slices}. We find a similar but less significant trend between the width of the \chieff distribution and the primary mass, where $\delta\hspace{-0.5mm}\log\sigma_{m_{1}}=0.09^{+0.10}_{-0.08}$, and $\delta\hspace{-0.5mm}\log\sigma_{m_1}=0$ is excluded at 93.5\% credibility. While the \chieff distributions in Fig.~\ref{fig:primary_mass_slices} appear visually to increase both in mean and width with increasing primary mass, the posterior on $\delta\mu_{m_1}$ is still consistent with 0 at 31.1\% credibility. In Section~\ref{sec:validation}, we comment on the possibility of a correlation between primary mass and \chieff being spuriously introduced due to mismodeling a true correlation between redshift and \chieff.

\begin{figure*}
\centering
\includegraphics[width=0.7\textwidth]{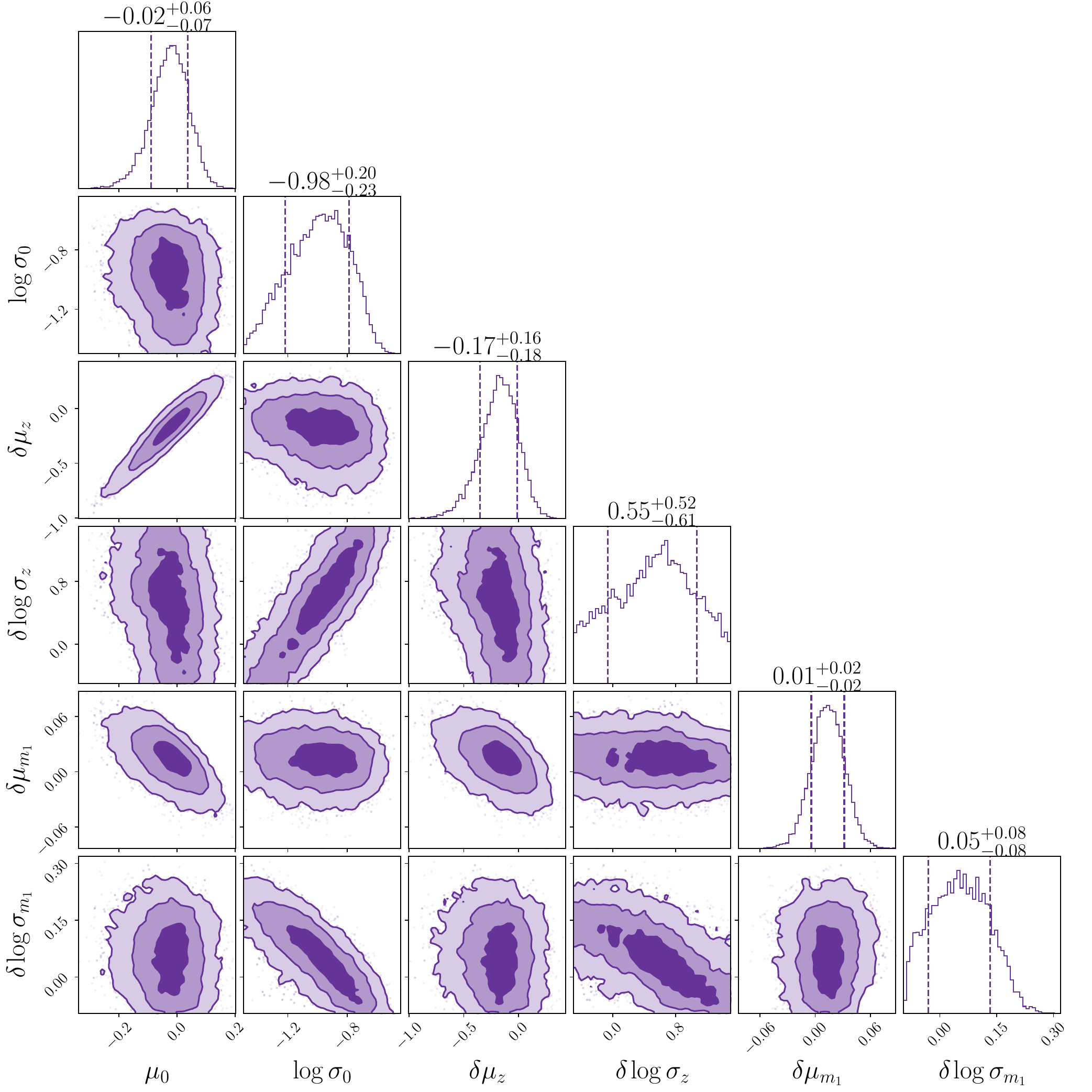}
\caption{Spin hyper-parameter posteriors obtained for GWTC-3 events when allowing for a correlation between \chieff and both primary mass and redshift. The colors indicate the 1, 2, and $3\sigma$ 2D credible regions, the dashed lines on the individual histograms show the 1D $1\sigma$ credible interval, and the median and $1\sigma$ credible interval are printed above each histogram.}
\label{fig:both_corner}
\end{figure*}

In order to determine if there is a preference in whether the redshift or the primary mass drives the evolution of the width of the \chieff distribution, we now analyze the data with a model that allows for correlations with both parameters. The posteriors on the spin hyper-parameters are shown in Fig.~\ref{fig:both_corner}. We obtain much weaker constraints on $\delta\hspace{-0.5mm}\log\sigma_{z}$ and $\delta\hspace{-0.5mm}\log\sigma_{m_1}$ individually, but both posteriors have more support for positive than negative values, and the point $\delta\hspace{-0.5mm}\log\sigma_{z}, \delta\hspace{-0.5mm}\log\sigma_{m_1} = (0,0)$ is disfavored, lying at the 96.2\% credibility contour. 

Based on these posteriors alone, we cannot confidently identify whether the correlation is dominated by the redshift or the primary mass, but comparing the Bayes factors between the various models that we have considered can provide an indication of which correlation is statistically preferred. Table~\ref{tab:bayes_factors} shows the natural log Bayes factors between the three correlated models we have presented so far and an uncorrelated model, where only $\mu_0$ and $\log\sigma_0$ in Eq.~\ref{eq:chi_eff_pop} are left as free parameters. We repeat the analysis allowing only for redshift correlations with fixed $\delta\mu_z=0$ in order to gauge the effect of the Occam penalty on the Bayes factor, since this sub-hypothesis is consistent with our initial redshift correlation result. While none of the values are particularly statistically significant, we find that the models including a correlation with primary mass are disfavored relative to the redshift-only models. We emphasize that while we have ensured consistent priors between all the models which are sub-hypotheses of each other, the choice of priors for individual hyper-parameters is necessarily somewhat arbitrary, complicating the interpretation of the Bayes factors\footnote{We can use a simple back-of-the-envelope calculation to determine the effect of the narrower width of the prior on $\delta\hspace{-0.5mm}\log\sigma_z$ on the Bayes factor for the redshift correlation models. Since the prior volume for $\delta\hspace{-0.5mm}\log\sigma_z$ is 2/3.5 times the prior volume for $\delta\hspace{-0.5mm}\log\sigma_q$ and $\delta\hspace{-0.5mm}\log\sigma_{m_1}$, the Bayes factor would be increased for the redshift correlation model by $\sim-\ln{2/3.5}\sim0.56$. This is comparable to the uncertainty in the Bayes factor due to the Monte Carlo integration in the likelihood in Eq.~\ref{eq:hyper-likelihood}.}.

\begin{table}
    \centering
\begin{tabular}{|p{4cm} ||p{1cm}|}
    \hline
    Correlation & $\ln(\mathrm{BF})$\\
    \hline
    Redshift, $\delta\mu_z = 0$ & 1.43 \\
    Redshift & 0.24\\
    Primary mass & -2.09\\
    Mass ratio & 3.75\\
    Redshift \& primary mass & -3.63\\
    Redshift \& mass ratio & 3.20\\
    \hline
\end{tabular}
\caption{Natural log Bayes factors comparing the models allowing for correlations between redshift, primary mass, mass ratio, and $\chieff$ and the base model without any correlations. The redshift-only models include both $\delta\mu_z$ and $\delta\hspace{-0.5mm}\log\sigma_z$ as free parameters and only $\delta\hspace{-0.5mm}\log\sigma_z$ as a free parameter with $\delta\mu_z$ fixed to zero. The models including a correlation with primary mass are disfavored by the data.}
\label{tab:bayes_factors}
\end{table}

Finally, we want to ensure that the apparent correlation between redshift and \chieff is not falsely introduced by the previously-reported correlation between the mean of the \chieff distribution and mass ratio~\citep{Callister:2021fpo, LIGOScientific:2021psn}. We modify the model in Eq.~\ref{eq:chi_eff_pop} to allow for correlations with both redshift and mass ratio rather than redshift and primary mass:
\begin{align}
    \mu_{\chi}(z, q) &= \mu_{0} + \delta\mu_z (z - 0.5) + \delta\mu_{q}\left(q -0.7\right)\\
    \log_{10}{\sigma_{\chi}}(z, q) &= \log\sigma_0 + \delta\hspace{-0.5mm}\log\sigma_z (z - 0.5) + \delta\hspace{-0.5mm}\log\sigma_{q}\left(q -0.7\right).
\end{align}
The priors on $\delta\mu_q$ and $\delta\hspace{-0.5mm}\log\sigma_q$ are given in Table~\ref{tab:spin_hyper_param_priors}, and the Bayes factor between this model and the uncorrlated model is also shown in Table~\ref{tab:bayes_factors}. The results we obtain with this model are consistent with both the previously-reported mass ratio correlation and the redshift correlation presented earlier in this work. The $\delta\mu_q$ posterior peaks at negative values, indicating that the mean of the \chieff distribution shifts towards smaller values as the mass ratio becomes more equal. Simultaneously, the $\delta\hspace{-0.5mm}\log\sigma_z$ posterior peaks at positive values, similar to the posteriors shown in Figs.~\ref{fig:spin_corner_z}, \ref{fig:both_corner}. Thus, we conclude that the increase in the width of the \chieff distribution that we report here is independent of the previously-established correlation between \chieff and mass ratio.

\section{Validation of results}
\label{sec:validation}
The results presented so far suggest that for the 69 BBH events we analyze with $\mathrm{FAR}<1~\mathrm{yr}^{-1}$ detected up to the end of O3, the \chieff distribution broadens with increasing redshift. As noted above, though, it is difficult to disentangle a correlation between spin and redshift from a correlation between spin and mass and to distinguish the extent to which the broadening of the \chieff distribution is due to increased uncertainty in spin measurements at high redshifts. We test the robustness of the observed redshift correlation first with a series of simulated populations, then by analyzing the GWTC-3 data with alternative models allowing for a correlation between \chieff and redshift. 

\subsection{Simulated populations}
One potential concern is that such a spin-redshift correlation could be introduced due to the degradation of the constraint on \chieff for individual events at farther redshifts, as these sources are generally detected with smaller signal-to-noise ratios (SNRs). To verify whether such a spurious correlation could be introduced, we simulate a population of 69 BBH events detected by a Hanford-Livingston detector network operating at O3 sensitivity~\citep{O3_psds} drawn from an intrinsically uncorrelated population. The true values of the hyper-parameters describing the population are given in Table~\ref{tab:injections}. We consider an event to be ``detected'' if it has a network optimal $\mathrm{SNR}\geq 9$. We perform individual-event parameter estimation using the reduced order quadrature implementation~\citep{Smith:2016qas} of the IMRPhenomPv2 waveform~\citep{Husa:2015iqa, Khan:2015jqa, Hannam:2013oca} and the \textsc{dynesty} nested sampler~\citep{Speagle_2020} through the \textsc{bilby} package~\citep{Ashton:2018jfp, Romero-Shaw:2020owr}. We then recover posteriors on the mass, redshift, and spin hyper-parameters using the three population models applied to the real data in the previous section. 

\begin{table}
    \centering
\begin{tabular}{|p{1.5cm} ||p{1.5cm} p{2.5cm}|}
    \hline
    Parameter & Value & Recovery\\
    \hline
    $\alpha$ & 4 & $4.74^{+1.33}_{-0.98}$\\
    $\beta$  & 1.5 & $1.02^{+2.10}_{-1.44}$ \\
    $m_{\max}$ & $50~\mathrm{M}_{\odot}$ & $46.63^{+46.35}_{-7.87}~\mathrm{M}_{\odot}$\\
    $m_{\min}$ & $5~\mathrm{M}_{\odot}$ & $4.99^{+0.87}_{-1.10}~\mathrm{M}_{\odot}$\\
    $\delta_{m}$ & $5~\mathrm{M}_{\odot}$ & $4.88^{+4.12}_{-2.12}~\mathrm{M}_{\odot}$\\
    $\mu_{m}$ & $35~\mathrm{M}_{\odot}$ & $31.40^{+1.59}_{-1.59}~\mathrm{M}_{\odot}$ \\
    $\sigma_{m}$ & $4~\mathrm{M}_{\odot}$ & $3.09^{+0.99}_{-2.09}~\mathrm{M}_{\odot}$ \\
    $\lambda$ & 0.04 & $0.012^{+0.019}_{-0.012}$\\
    $\lambda_{z}$ & 3 & $5.15^{+1.83}_{-1.83}$\\
    $\mu_{0}$ & 0.05 & $0.025^{+0.076}_{-0.088}$\\
    $\log\sigma_{0}$ & -0.85 & $-0.82^{+0.22}_{-0.20}$\\
    $\delta\mu_{z}$ & 0 & $-0.17^{+0.21}_{-0.11}$\\
    $\delta\hspace{-0.5mm}\log\sigma_{z}$ & 0 & $-0.24^{+0.62}_{-0.47}$\\
    $\delta\mu_{m_1}$ & 0 & $0.023^{+0.035}_{-0.035}$\\
    $\delta\hspace{-0.5mm}\log\sigma_{m_1}$ & 0 & $-0.14^{+0.07}_{-0.08}$\\
    \hline
\end{tabular}
\caption{True and recovered values of the hyper-parameters describing the mass, spin, and redshift distributions from which the simulated population with no correlation was drawn. The correlated population is described by the same hyper-parameters with the exception of $\delta\hspace{-0.5mm}\log\sigma_{z}=0.85$. The recovered values are represented by the maximum posterior value and 90\% credible interval calculated with the maximum posterior density method obtained with the model that allows for both primary mass and redshift correlations.}
\label{tab:injections}
\end{table}

The true values of all hyper-parameters describing this simulated population are recovered within at least the $3\sigma$ level, although the posteriors for both $\delta\hspace{-0.5mm}\log\sigma_{z}$ and $\delta\hspace{-0.5mm}\log\sigma_{m_1}$ tend to prefer negative values. This suggests that it is easier to rule out a distribution that broadens with redshift or mass rather than one that becomes narrower. A broadening distribution implies the presence of highly spinning sources at high mass/redshift, which are easier to detect. Therefore, their absence from the observed population indicates that they are also absent from the underlying population. The converse is not true. If the \chieff distribution instead became narrower, there would be an increasing number of sources with nearly zero spin at high mass/redshift. These sources are harder to detect, so their absence from the observed distribution does not necessarily imply their absence from the underlying distribution. 

\begin{figure}
\centering
\includegraphics[width=\columnwidth]{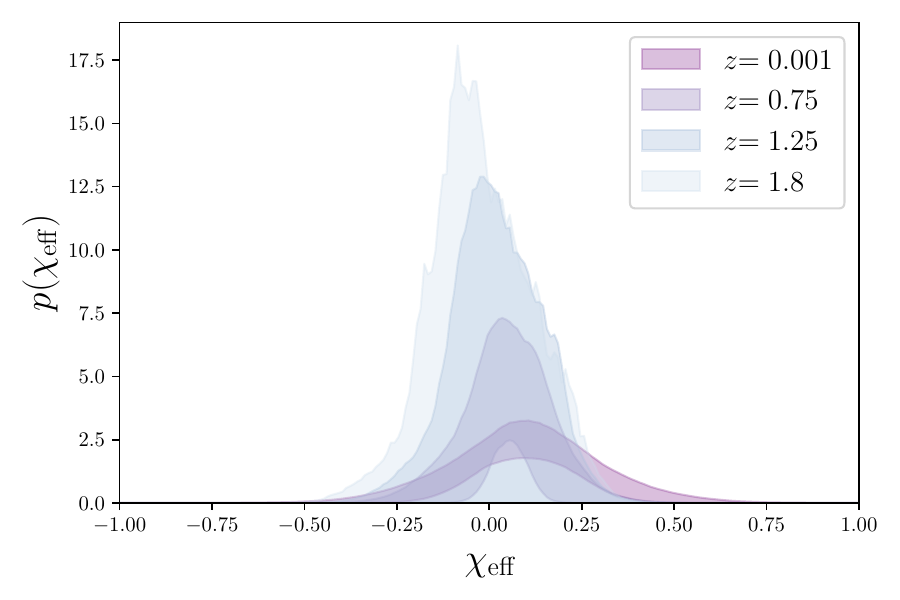}
\includegraphics[width=\columnwidth]{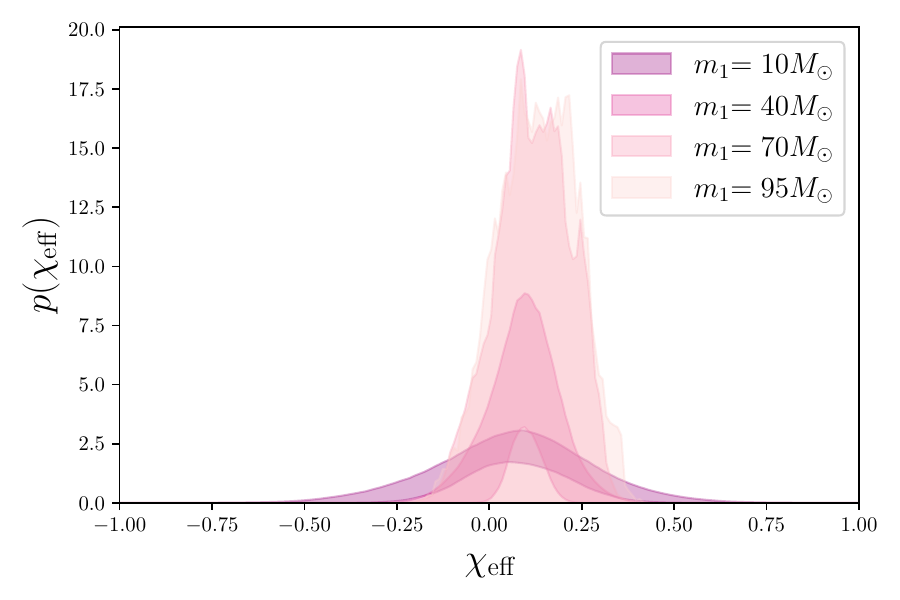}
\caption{90\% credible region for slices of the \chieff distribution at four different values of redshift (top) and primary mass (bottom) for the simulated population with no correlations.
}
\label{fig:chieff_slice_inj}
\end{figure}

The \chieff distributions along different slices in redshift and primary mass are shown in Fig.~\ref{fig:chieff_slice_inj}. These distributions are morphologically distinct from those obtained for the real data in Figs.~\ref{fig:slices_z} and \ref{fig:primary_mass_slices}. The distributions obtained for this uncorrelated population appear to initially get narrower between the lower two values of redshift and primary mass but then broaden again. This indicates that there is no strong evidence for either an increase or a decrease in the width of the \chieff distribution with either parameter. There is more uncertainty in the distributions at higher primary mass and redshift, a feature which is not observed for the real data, further increasing our confidence in our measurement of a broadening in the spin distribution with mass and/or redshift.

The natural log Bayes factors between each of the models allowing for correlations and the model without any correlations for this simulated population are given in Table~\ref{tab:bayes_factors_inj_uncorrelated}. In this case, the redshift-correlation model is disfavored at a level comparable to the primary-mass-correlation model, as expected since the true population does not contain any correlations. This simulation suggests that a positive correlation between the width of the \chieff distribution and either the redshift or primary mass is unlikely to be spuriously introduced by the worsening constraint on \chieff for individual events at higher redshift or primary mass.

\begin{table}
    \centering
\begin{tabular}{|p{3cm} ||p{1cm}|}
    \hline
    Correlation & $\ln(\mathrm{BF})$\\
    \hline
    Redshift & -1.93\\
    Primary mass & -1.59\\
    Both & -3.35\\
    \hline
\end{tabular}
\caption{Natural log Bayes factors comparing the models allowing for correlations between redshift, primary mass, and $\chieff$ and the base model without any correlations for the simulated population with no correlations.}
\label{tab:bayes_factors_inj_uncorrelated}
\end{table}

\begin{figure}
\centering
\includegraphics[width=\columnwidth]{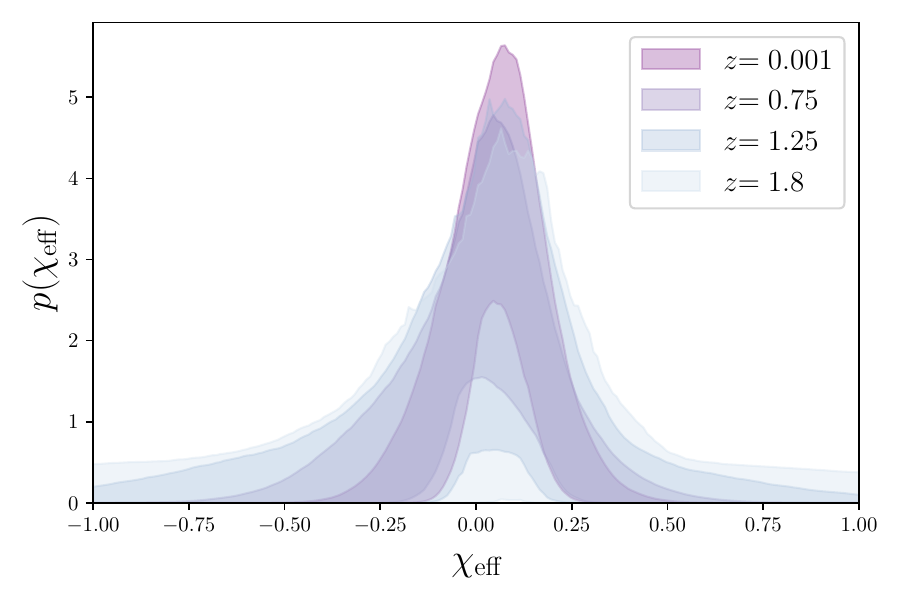}
\includegraphics[width=\columnwidth]{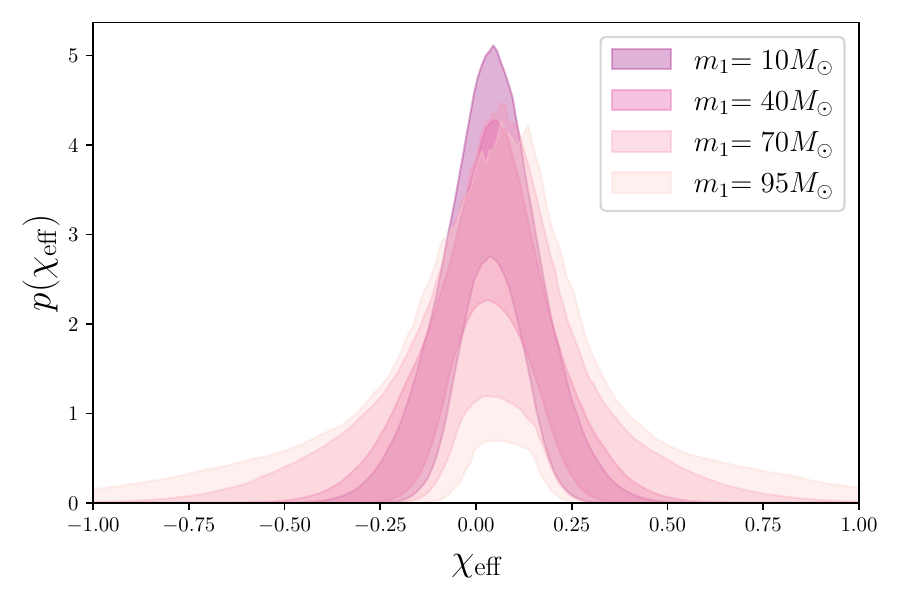}
\caption{90\% credible region for slices of the \chieff distribution at four different values of redshift (top) and primary mass (bottom) for the simulated population with a correlation between the width of the \chieff distribution and redshift.}
\label{fig:slices_inj_correlated}
\end{figure}

We next seek to verify if we are able to successfully recover a correlation in a simulated population similar to the one we find in real data, and whether a redshift correlation can manifest as a primary mass correlation when analyzed with the wrong model. We again generate 69 sources detected at O3 sensitivity now drawn from a distribution with $\delta\hspace{-0.5mm}\log\sigma_{z}=0.85$. All the other true hyper-parameter values are the same as for the uncorrelated population given in Table~\ref{tab:injections}. We initially analyze this simulated population allowing for mass and redshift correlations in turn. As before, all the hyper-parameter values are recovered within $3\sigma$ credibility, although the posterior on $\delta\hspace{-0.5mm}\log\sigma_{z}$ peaks below the true value. The distributions for \chieff along different slices in mass and redshift shown in Fig.~\ref{fig:slices_inj_correlated} are similar to those recovered in the real data. They become consistently wider between increasing values of redshift and primary mass, although they also become more uncertain, which is not observed in the real data. This is because the posteriors for individual events in the simulated data set only extend up to $z=1.14, \ m_{1}=104~M_{\odot}$, while those for real events include values up to $z=1.90,\ m_{1}=296~M_{\odot}$, meaning there is more resolving power for the mass and redshift distributions at large masses and redshifts in the real data.

While the posterior on $\delta\hspace{-0.5mm}\log\sigma_{m_1}$ includes 0 within the 81.4\% credible level, this simulation indicates that a true correlation between redshift and \chieff could be perceived as a correlation between primary mass and \chieff if analyzed with the wrong model. It also reinforces the significance of our finding a preference for positive values of $\delta\hspace{-0.5mm}\log\sigma_z$ in the real data, since both simulations tend to recover smaller values of $\delta\hspace{-0.5mm}\log\sigma_z$ compared to the true value. This is further supported by the Bayes factors we recover for the correlated simulation, given in Table~\ref{tab:bayes_factors_inj}. In this case, the recovered correlation with redshift-only is not significant enough to overcome the Occam penalty for adding parameters relative to the uncorrelated model. However, the relative Bayes factors between the three models considered are similar to those obtained for the real data. This indicates that the redshift-only correlation found in the real data is preferred over the other correlated models we explore at a similar level of statistical significance to the simulated population with a known correlation in $\delta\hspace{-0.5mm}\log\sigma_z$. On the other hand, the relative Bayes factors for the uncorrelated simulation span a much narrower range, revealing that none of the correlated models is strongly preferred over the others in the case of no correlation.

\begin{table}
    \centering
\begin{tabular}{|p{3cm} ||p{1cm}|}
    \hline
    Correlation & $\ln(\mathrm{BF})$\\
    \hline
    Redshift & -2.45\\
    Primary mass & -4.55\\
    Both & -6.97\\
    \hline
\end{tabular}
\caption{Natural log Bayes factors comparing the models allowing for correlations between redshift, primary mass, and $\chieff$ and the base model without any correlations for the simulated population with $\delta\hspace{-0.5mm}\log\sigma_z=0.85$.}
\label{tab:bayes_factors_inj}
\end{table}

Conversely, it is also possible that a true correlation between primary mass and \chieff could manifest as a correlation between redshift and \chieff if analyzed with the wrong model. To verify whether this is a potential false source of the redshift-\chieff correlation we observe, we simulate a third population of 69 events detected at O3 sensitivity with hyper-parameters identical to the previous two, except now $\delta\hspace{-0.5mm}\log{\sigma_z}=0$ and $\delta\hspace{-0.5mm}\log{\sigma_{m_1}}=0.15$. We find that when this population is analyzed with the model allowing for a redshift correlation only, $\delta\hspace{-0.5mm}\log{\sigma_{z}}=0$ is incorrectly ruled out at 99.0\% credibility, indicating that a primary mass-\chieff correlation can indeed be confused for a redshift-\chieff correlation. This is similar to the case explored in the previous two paragraphs where $\delta\hspace{-0.5mm}\log{\sigma_{m_1}}=0$ was disfavored at the 81.4\% credible level for a true correlation between redshift and spin analyzed with the wrong model.

\begin{figure}
\centering
\includegraphics[width=\columnwidth]{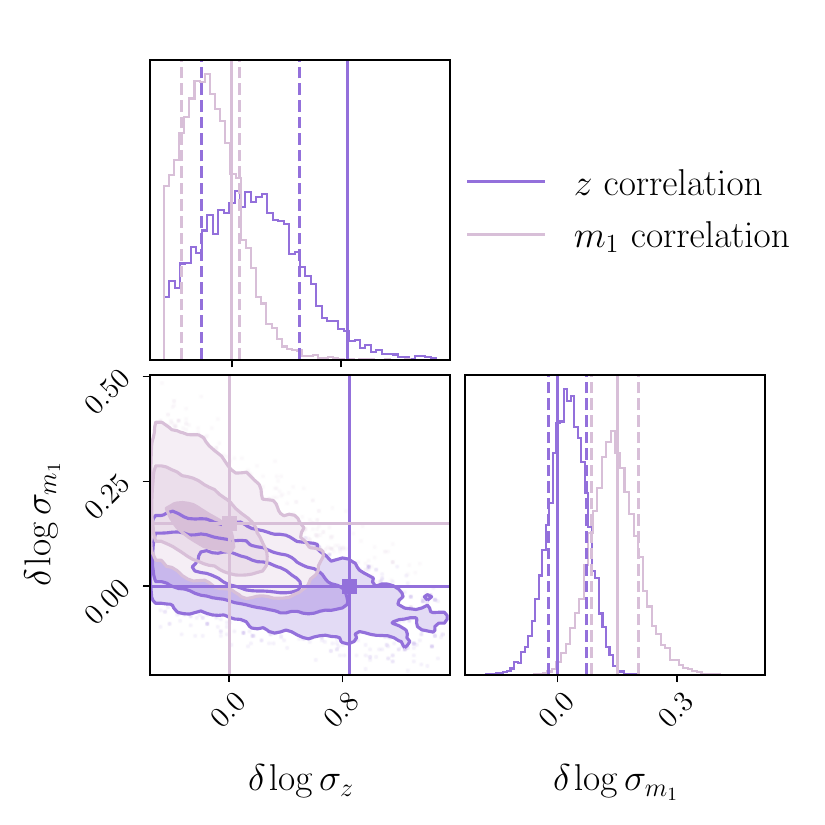}
\caption{Posteriors for $\delta\hspace{-0.5mm}\log{\sigma_{m_1}}, \delta\hspace{-0.5mm}\log{\sigma_{z}}$ obtained under the model allowing for both redshift and primary mass correlations with \chieff, for two different simulated populations. The dark purple shows the results for the population with a simulated redshift correlation, $\delta\hspace{-0.5mm}\log{\sigma_{z}}=0.85$, while the light purple shows the results for the population with a simulated primary mass correlation, $\delta\hspace{-0.5mm}\log{\sigma_{m_1}}=0.15$. The shading indicates the 1, 2, and $3\sigma$ 2D credible regions and the dashed lines on the individual histograms show the 1D $1\sigma$ credible interval. While the posterior for the redshift-correlated population is consistent with primary mass being the sole source of correlation, the opposite is not true.}
\label{fig:inj_both_corner_comp}
\end{figure}

However, we find in general that a correlation between spin and primary mass is easier to confidently identify with 69 events detected at O3 sensitivity. The posterior on $\delta\hspace{-0.5mm}\log{\sigma_{m_1}}$ disfavors 0 at 99.7\% credibility under the model allowing only for a primary mass correlation and at 99.5\% credibility under the model allowing for both primary mass and redshift correlations. This is illustrated in the corner plot of the posteriors on $\delta\hspace{-0.5mm}\log{\sigma_{m_1}}, \delta\hspace{-0.5mm}\log{\sigma_{z}}$ in Fig.~\ref{fig:inj_both_corner_comp} for both the mass-correlated population and the redshift-correlated population previously presented in Fig.~\ref{fig:slices_inj_correlated}. While it is difficult to experimentally distinguish between mass and redshift evolution of the spin distribution in the case of a redshift-correlated population, the same ambiguity is not present for the mass-correlated population. This further supports the redshift-\chieff correlation we find in the real data, since a correlation between primary mass and \chieff would likely have been unambiguously identified. 

\subsection{Alternative models}
\label{sec:alternatives}
While the results of the two simulated populations lend credence to the finding that the width of the \chieff distribution increases with redshift in the real data, we want to verify whether this result is model-driven. To this end, we now analyze the GWTC-3 events with a different model---a mixture of truncated Gaussians with a redshift-dependent mixing fraction,
\begin{align}
\pi(\chieff | \Lambda_{\chieff}) &= \mathcal{N}(\chieff; \mu_b, 10^{\sigma_b})(1 - f(z)) \\&+ \mathcal{N}(\chieff; \mu_a, 10^{\sigma_a})f(z),
\end{align}
where
\begin{align}
f(z) &= \frac{1}{1 + \exp{\left[f_{A} + f_{B}(z - 0.5)\right]}}.
\label{eq:double}
\end{align}
We choose three different priors on the $\mu$ and $\sigma$ parameters for each Gaussian, outlined in Table~\ref{tab:spin_double_hyper_param_priors}, as proxies for different astrophysical scenarios. 

\begin{table*}
    \centering
\begin{tabular}{|p{1.5cm} ||p{6cm} p{3.5cm} p{2cm} p{2cm}|}
    \hline
    Parameter & Description & Prior 1 & Prior 2 & Prior 3\\
    \hline
    $\mu_{a}$ & Mean of the secondary Gaussian & $\mathrm{U}(-1, -0.1) \cup \mathrm{U}(0.1, 1)$ & $\mathrm{U}(-1, 1)$ & $\mathrm{U}(0.1, 1)$ \\
    $\mu_{b}$ & Mean of the bulk Gaussian & 0.06 & 0 & 0\\
    $\sigma_{a}$ & Log-width of secondary Gaussian & $\mathrm{U}(-1.5, 0.5)$ & $\mathrm{U}(-0.7, 1.5)$ & $\mathrm{U}(-1.5, 0.5)$\\
    $\sigma_{b}$ & Log-width of the bulk Gaussian & $\mathrm{U}(-1.5, 0.5)$ & $\mathrm{U}(-2, -0.7)$ & $\mathrm{U}(-1.5, 0.5)$\\
    $\chi_{a, \min}$ & Lower bound of the secondary Gaussian & -1 & -1 & 0\\ 
    $f_{A}$ & Independent offset in mixing fraction & $\mathcal{N}(0, 1.5)$ & $\mathcal{N}(0, 1.5)$ & $\mathcal{N}(0, 1.5)$\\
    $f_{B}$ & $z$-dependent offset in mixing fraction & $\mathcal{N}(0, 1.5)$ & $\mathcal{N}(0, 1.5)$ & $\mathcal{N}(0, 1.5)$\\
    \hline
\end{tabular}
\caption{Spin hyper-parameter priors and descriptions for the model in  Eq.~\ref{eq:double}.}
\label{tab:spin_double_hyper_param_priors}
\end{table*}

\begin{figure}
\centering
\includegraphics[width=\columnwidth]{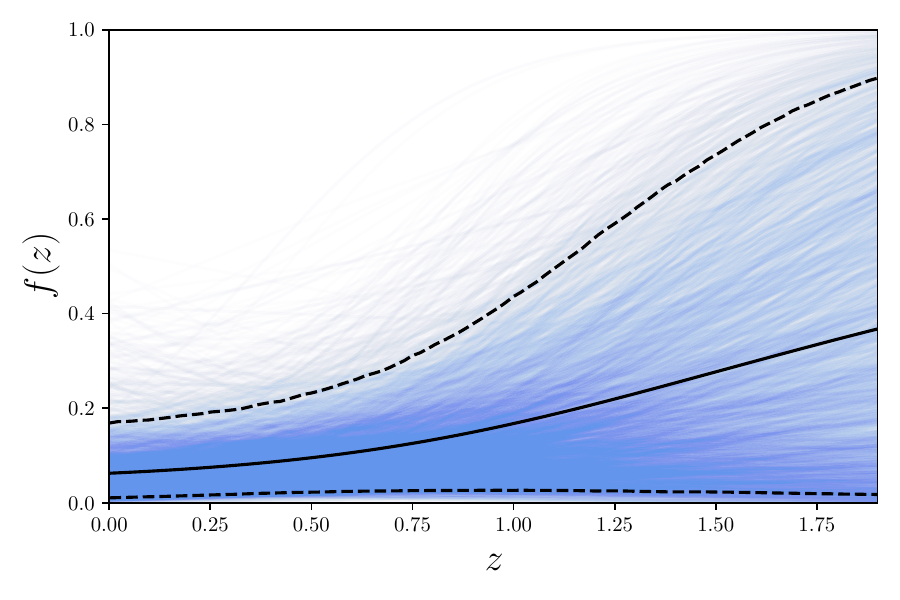}
\caption{Posteriors for the mixture fraction in Eq.~\ref{eq:double} obtained for GWTC-3 events under the first prior choice in Table~\ref{tab:spin_double_hyper_param_priors}. The solid black line shows the mean, while the dashed black lines show the 90\% credible region.}
\label{fig:frac_scaling}
\end{figure}

\begin{figure}
\centering
\includegraphics[width=\columnwidth]{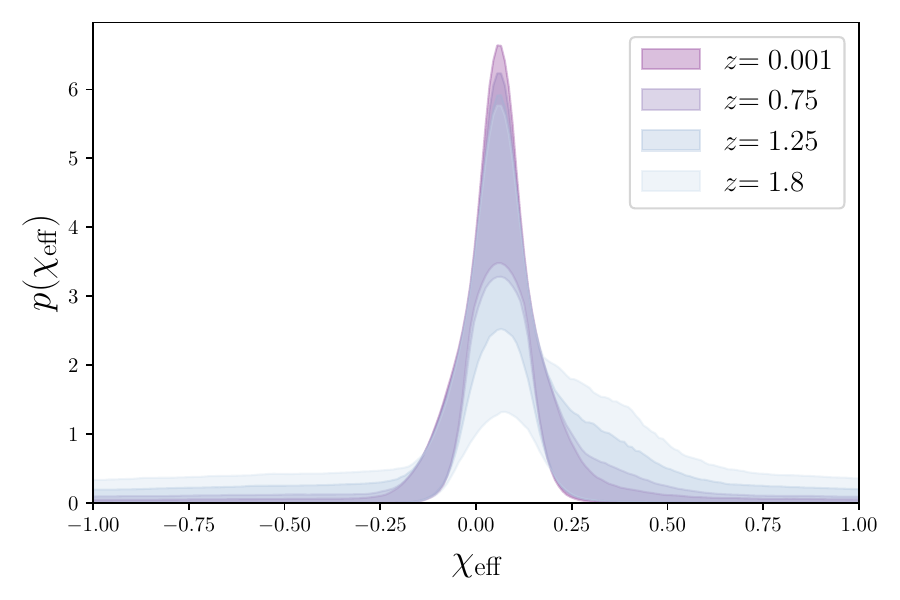}
\includegraphics[width=\columnwidth]{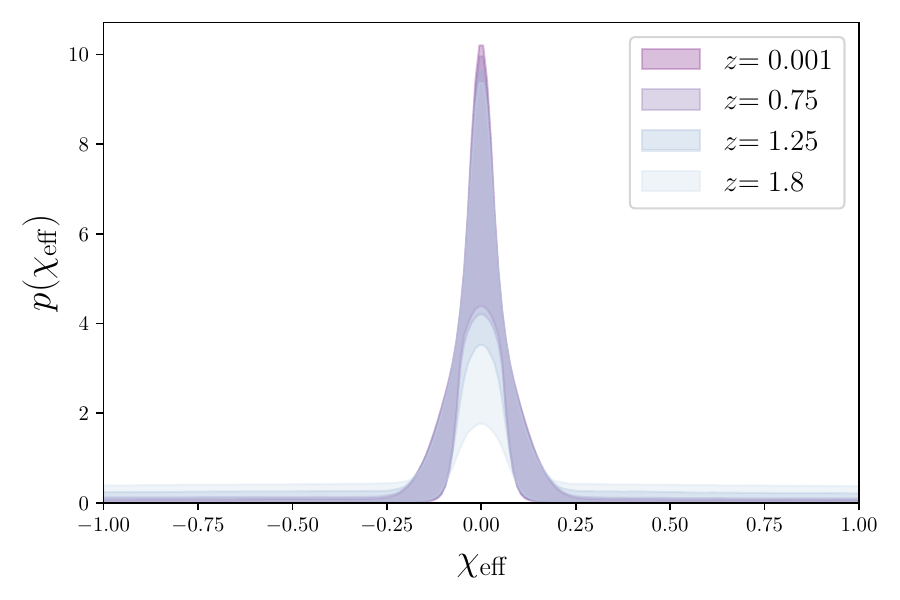}
\includegraphics[width=\columnwidth]{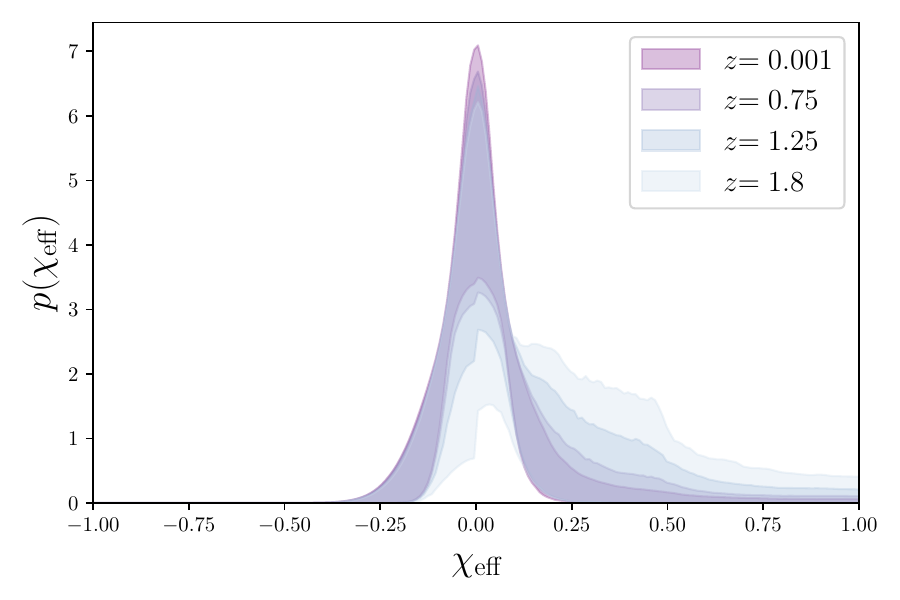}
\caption{90\% credible region for slices of the \chieff distribution at four different values of redshift for the three different prior choices in Table~\ref{tab:spin_double_hyper_param_priors} (1-3 from top to bottom).}
\label{fig:redshift_mix_slices}
\end{figure}

When modeling the effective spin distribution with a single, redshift-independent Gaussian, it is found that the mean effective spin is $\chieff=0.06$~\citep{Miller:2020zox, LIGOScientific:2021psn}. For our first prior choice, we assume that this result encapsulates the bulk of the population, fixing $\mu_b=0.06$, and let the second mixture component model the departure from this assumption as we look to higher redshifts. We allow this secondary sub-population to peak at either positive or negative values of \chieff, but the absolute value of the peak must be $\geq 0.1$, so as to minimize the degeneracy between the bulk and this second sub-population. The evolution of the mixture fraction with redshift for individual hyper-parameter posterior samples is shown in Fig.~\ref{fig:frac_scaling}. The posterior on the mixture fraction differs substantially from the prior, which is symmetric about $f=0.5$ for all redshifts. This result indicates that the contribution of the Gaussian defined by the parameters $\mu_a$ and $\sigma_a$ becomes more significant as redshift increases. The posterior for $\mu_a$ is 3.4 times more likely to be positive than negative, and the positive part of the posterior rails against the boundary at $\mu_a=0.1$, indicating that it may be trying to replicate the bulk distribution rather than to extract an independent component. $\sigma_a$ is only weakly constrained to be $>-1.02$ at 90\% credibility. Nonetheless, the \chieff distributions at different redshift slices shown in the top of Fig.~\ref{fig:redshift_mix_slices} for this model and prior choice paint a similar picture to the linear evolution model presented in Section~\ref{sec:results}; although skewed towards positive values of \chieff, the distributions become wider with increasing redshift.

The second prior choice targets two populations with \chieff distributions characterized by different widths. The peak of the narrower bulk distribution is now fixed to $\mu_b=0$ to capture the BBH population formed dynamically with small spin and the majority of field binaries predicted to have negligible spin~\citep{Fuller:2019sxi}. Meanwhile the posterior on the peak of the broader Gaussian, $\mu_a$, is largely uninformative. The posteriors for the width parameters and the mixture fraction are very similar to those obtained with the first prior, with $\sigma_a > -0.25$ at 90\% credibility and $\sigma_b$ more strongly constrained to peak at $-1.23^{+0.17}_{-0.17}$. These results do not provide significant evidence for two distinct populations, but the \chieff distributions at different redshift slices shown in the middle panel of Fig.~\ref{fig:redshift_mix_slices} still exhibit some broadening with increasing redshift.

The final prior choice is designed to look for a redshift-dependent excess of preferentially-aligned BBH systems. The mean of the bulk distribution is still fixed to $\mu_b=0$, as expected for a population of dynamically-formed systems with isotropic spin tilts~\citep{PortegiesZwart:2002iks,Antonini:2016gqe, Rodriguez:2015oxa, Gerosa:2017kvu, Rodriguez:2019huv, Kimball:2020opk, Gerosa:2021mno}. The mean of the aligned population is restricted to $0.1<\mu_a<1$, in order to represent systems formed via isolated binary evolution in the field~\citep{Gerosa:2018wbw,Qin:2018vaa,Zaldarriaga:2017qkw,Belczynski:2017gds,Bavera:2020inc,Bavera:2020uch, Bavera:2022mef}, and the corresponding Gaussian is truncated on $[0,1]$ instead of $[-1,1]$. The \chieff distributions at different redshift slices for this prior choice are shown in the bottom panel of Fig.~\ref{fig:redshift_mix_slices}. The population model is asymmetric by construction, so the mean of the distribution also increases with redshift along with the width.

The posterior on the mixture fraction is again very similar to the one obtained with the first prior choice in Fig.~\ref{fig:frac_scaling}. Since the mixture fraction increases with redshift for all three prior choices, the second Gaussian added on top of the bulk distribution always contributes more at high redshifts, consistent with the conclusion that the \chieff distribution broadens with increasing redshift. These results indicate that the apparent correlation between the width of the \chieff distribution and the redshift is not driven by our choice of linear evolution model in Section~\ref{sec:methods}, but can be observed under a variety of population models and priors.

\section{Discussion and Conclusion}
\label{sec:conclusions}
In this work, we have found weak but robust evidence for an increase in the width of the \chieff distribution of BBH systems with increasing redshift. We have verified that this correlation is unlikely to be spuriously introduced by the increased uncertainty in the \chieff posteriors for individual high-redshift sources using a simulated population and that this trend remains present using alternative population models. We also observe a less significant correlation between the width of the \chieff distribution and the primary mass, although we find that such a correlation can be falsely recovered if a population with a \chieff-redshift correlation is instead analyzed assuming only a \chieff-primary mass correlation. When allowing for correlations with both redshift and primary mass, we find evidence for an increase in the width of the \chieff distribution with one or the other but cannot distinguish which.

A correlation between $\chieff$ and redshift might be explained by one of two broad scenarios.
First, the correlation could be indicative of multiple sub-populations arising from distinct formation channels, each of which occupy a different region in the $(\chieff,z)$ plane.
The third model defined in Section~\ref{sec:alternatives}, for example, serves as a proxy for this scenario.
In this case, the symmetric $\chieff$ component centered at zero might serve to capture binaries assembled dynamically in dense stellar environments~\citep{PortegiesZwart:2002iks,Antonini:2016gqe, Rodriguez:2015oxa, Gerosa:2017kvu, Rodriguez:2019huv, Kimball:2020opk, Gerosa:2021mno}, while the preferentially-positive $\chieff$ component corresponds to a sub-population of systems formed via isolated binary evolution in the field~\citep{Gerosa:2018wbw,Qin:2018vaa,Zaldarriaga:2017qkw,Belczynski:2017gds,Bavera:2020inc,Bavera:2020uch, Bavera:2022mef}.
However, such a mixture between two sub-populations generally leads to a shift in the \textit{mean} of the $\chieff$ distribution with redshift.
Under the more generic model explored in Section~\ref{sec:results}, we much more strongly prefer evolution of the \textit{width} with redshift, although evolution of $\mu_\chi$ is not strictly ruled out.

The second possibility, broadly, is that a spin-redshift correlation arises due to evolutionary processes operating within a single population.
Within isolated binary evolution, for example, tidal interactions may be responsible for correlating black hole spins and redshifts.
Some authors predict that isolated black holes are naturally born slowly rotating due to efficient angular momentum transport from stellar cores~\citep{Spruit:2001tz,2019MNRAS.485.3661F,Fuller:2019sxi}.
Spin can nevertheless be introduced via tidal torques exerted by the first-born black hole on its companion, prior to the companion's core collapse~\citep{Zaldarriaga:2017qkw,Qin:2018vaa,Bavera:2020uch, Bavera:2020inc,Fuller:2022ysb}.
In this scenario, the shortest-period binaries both acquire the largest spins and merge most promptly, correlating the spins of binary black holes and the redshifts at which they merge~\citep{Qin:2018vaa,Fuller:2019sxi}.
This effect may be enhanced by the lower metallicities present at high redshifts, which diminish the strength of stellar winds and hence prevent binary orbits from widening and avoiding tidal spin-up.
\cite{Bavera:2022mef} corroborate this prediction using population synthesis simulations, finding that an increasing fraction of systems undergo tidal spin-up in close orbits at high redshift. Because there are systems that do not meet the criteria for efficient tidal spin-up at all redshifts, they find that the spin distribution both broadens and increases in mean.

Another possible explanation comes from the effect of metallicity on the efficiency of angular momentum transport in the envelope of the stellar progenitors of the BBH system. Because stellar winds are weaker at low metallicity (higher redshift), this leads to a less efficient removal of the angular momentum stored in the stellar envelope and hence a possibly higher spin of the resulting black hole~\citep{Qin:2018vaa,Fuller:2019sxi}. 
However, this naive picture is complicated by the interplay between the strength of stellar winds, the extent of the hydrogen-burning region inside the star, and the efficiency of elemental mixing within the star. As such, \cite{Belczynski:2017gds} find a non-monotonic relationship between the metallicity and the black hole spin for a given carbon-oxygen-core mass of the progenitor. This relationship further depends on the stellar evolution model employed in the binary evolution simulation. Because of the uncertainty in the processes that impart natal spin to the black hole, we cannot place meaningful constraints on these models.

While both of the possibilities described above lead to systems with larger spin magnitudes at increasing redshift, we must also explain the increased number of systems with negative values of \chieff due to the symmetric broadening of the distribution.
One potential explanation within the framework of isolated binary evolution is that natal kicks are stronger at higher redshift, leading to more significant misalignment of the BH spin relative to the orbital angular momentum upon birth.
It is not clear how such an effect would arise, however.
Furthermore, even with moderately strong ($\sim 100\mathrm{km}\,\mathrm{s}^{-1}$) natal kicks, the median of the \chieff distribution for systems formed in the field is still much higher than what we recover~\citep{Gerosa:2018wbw}. Alternatively, the mixing fraction between systems formed via isolated evolution and dynamical assembly may remain constant with redshift, but the spin \textit{magnitudes} of systems formed via both channels can increase due to one or both of the scenarios previously discussed. In this case, the orientations of the spins in systems formed dynamically would remain isotropic, but their magnitudes would be larger at higher redshift, leading to a broadening of the distribution but not a shift in the mean.

One caveat with our analysis is that we restrict any correlations to be with \chieff, rather than also allowing for mass-redshift correlations. It is likely that the complex processes leading to the formation of BBH systems would produce correlations between all of these parameters, rather than just with spin~\citep{Qin:2018vaa,Bavera:2020uch,Fuller:2022ysb,Zevin:2022wrw}. However, it is difficult to encapsulate all possible correlations in a simple phenomenological model without dramatically increasing the dimensionality of the parameter space. We leave the exploration of simultaneous correlations between all parameters to future work.

\acknowledgments
The authors thank Thomas Dent and Colm Talbot for useful discussions and Michael Zevin, Maya Fishbach, and Jos\'{e} Nu\~{n}o for helpful comments on the manuscript.
S.B., C-J.H., K.K.Y.N., and S.V. acknowledge support of the National Science Foundation and the LIGO Laboratory.

LIGO was constructed by the California Institute of Technology and Massachusetts Institute of Technology with funding from the National Science Foundation and operates under cooperative agreement PHY-0757058. This research has made use of data, software and/or web tools obtained from the Gravitational Wave Open Science Center (\url{https://www.gw-openscience.org}), a service of LIGO Laboratory, the LIGO Scientific Collaboration and the Virgo Collaboration.
S.~V. is supported by the NSF through award PHY-2045740.
S.~B. is also supported by the NSF Graduate Research Fellowship under grant No. DGE-1122374. The Flatiron Institute is a division of
the Simons Foundation, supported through the generosity of Marilyn and Jim Simons.
The authors are grateful for computational resources provided by the LIGO Lab and supported by NSF Grants PHY-0757058 and PHY-0823459.
This paper carries LIGO document number LIGO-P2200105.

\bibliography{spins}
\end{document}